# Enhancing Fully Formatted End-to-End Speech Recognition with Knowledge Distillation via Multi-Codebook Vector Quantization

Jian You
Cisco Systems
Shanghai, China
jianyou@cisco.com

Xiangfeng Li
Cisco Systems
Shanghai, China
xiangfel@cisco.com

Erwan Zerhouni
Cisco Systems
Zurich, Switzerland
ezerhoun@cisco.com

*Abstract*—Conventional automatic speech recognition (ASR) models typically produce outputs as normalized texts lacking punctuation and capitalization, necessitating post-processing models to enhance readability. This approach, however, introduces additional complexity and latency due to the cascaded system design. In response to this challenge, there is a growing trend to develop end-to-end (E2E) ASR models capable of directly predicting punctuation and capitalization, though this area remains underexplored. In this paper, we propose an enhanced fully formatted E2E ASR model that leverages knowledge distillation (KD) through multi-codebook vector quantization (MVQ). Experimental results demonstrate that our model significantly outperforms previous works in word error rate (WER) both with and without punctuation and capitalization, and in punctuation error rate (PER). Evaluations on the LibriSpeech-PC test-clean and test-other subsets show that our model achieves state-of-the-art results.

## I. INTRODUCTION

In conventional automatic speech recognition (ASR) models, output texts are often presented in lowercase and without punctuation, which significantly diminishes readability and user experience. The lack of proper formatting not only hampers direct usability but also adversely affects the performance of downstream applications like summarization, named entity recognition (NER) and neural machine translation (NMT), which rely on well-formatted text. Consequently, the inability of conventional ASR systems to produce fully formatted outputs undermines their utility in more sophisticated natural language processing tasks. This deficiency necessitates the employment of supplementary post-processing models to format the text.

Numerous studies have focused on the restoration of punctuation and capitalization in unformatted text. Early research predominantly utilized convolutional neural networks (CNNs) and recurrent neural networks (RNNs) for punctuation restoration [1, 2, 3]. With the advent and growing popularity of the transformer model [4], known for its superior performance and ability to capture contextual information over longer sequences compared to RNNs, many recent studies have adopted transformers for this task [5, 6, 7]. However, integrating a post-processing module for punctuation and capitalization prediction in conventional ASR systems poses several challenges. Primarily, these modules often neglect acoustic features like prosody, which are crucial for accurate orthographic judgments, such as distinguishing between statements and questions. Relying solely on lexical features increases vulnerability to errors and misinterpretations. Moreover, chaining separate models for formatting demands considerable maintenance and introduces latency, complicating the system's efficiency.

In view of the limitations inherent in cascaded systems, recent research has increasingly focused on integrating formatting functions into conventional ASR systems. This innovative end-to-end approach allows for the direct mapping of fully formatted text token sequences from acoustic features. E2E formatted speech recognition was initially explored in [8], where the system directly output case-sensitive text from speech input. Subsequent studies by [9] and [10] developed systems capable of generating punctuated text directly from speech in an end-to-end manner. In [11], stacked transformer encoder layers were employed and trained using connectionist temporal classification (CTC) loss with paired speech and punctuated text data. Similarly, [12] utilized a transformer model while introducing an independent decoder specifically for punctuation. Although the off-the-shelf transformer-transformer architecture known as Whisper [13] demonstrated that ASR models could be trained to produce formatted text, its inability to function in streaming mode restricts its applicability in real-time scenarios. Both [14] and [15] trained conformer-based [16] E2E models on texts with punctuation and capitalization, achieving commendable performance on public corpora. In studies [17] and [18], pretrained large language models (LLMs) were integrated with E2E ASR systems to achieve fully formatted transcription outputs.

Despite some studies on fully formatted E2E ASR systems, this field remains underexplored. In this study, we present an enhanced fully formatted E2E ASR model, which is based on knowledge distillation via multi-codebook vector quantization. Specifically, our approach employs a transformer-like transducer system, trained on formatted texts with recurrent neural network transducer (RNN-T) [19] loss. The recently published model Zipformer [20] is selected as the acoustic encoder for its speed and high performance. Inspired by prior work [11] where an auxiliary CTC loss is calculated from the output of the intermediate layer and the token sequence without punctuation, we adopt a knowledge distillation framework [21] to teach the intermediate layer of the acoustic encoder to learn from the embeddings at a different intermediate layer of a self-supervised, pre-trained teacher model. While knowledge distillation has been commonly used in traditional ASR systems [22, 23, 24], to the best of our knowledge, our research represents the first endeavor to enhance fully formatted ASR with knowledge distillation in an E2E manner. The motivation for incorporating knowledge distillation is to enable the intermediate layer of the acoustic encoder to more effectively facilitate accurate text prediction. Punctuation prediction, on one hand, is influenced by prosodic features; on the other hand, it is dependent on the context provided by previously predicted text, given that the RNN-T decoder conditions each new output on the sequence of previously generated non-blank labels. Hence, more accurate previously predicted text enhances the model's ability to accurately predict punctuation. Furthermore, as capitalization and punctuation prediction can benefit from each other [7], improvements in punctuation prediction can also yield enhancements in the accurate prediction of capitalization. Experiments conducted on the LibriSpeech-PC [15] datasets demonstrate the proposed model's superior performance, validating its effectiveness.

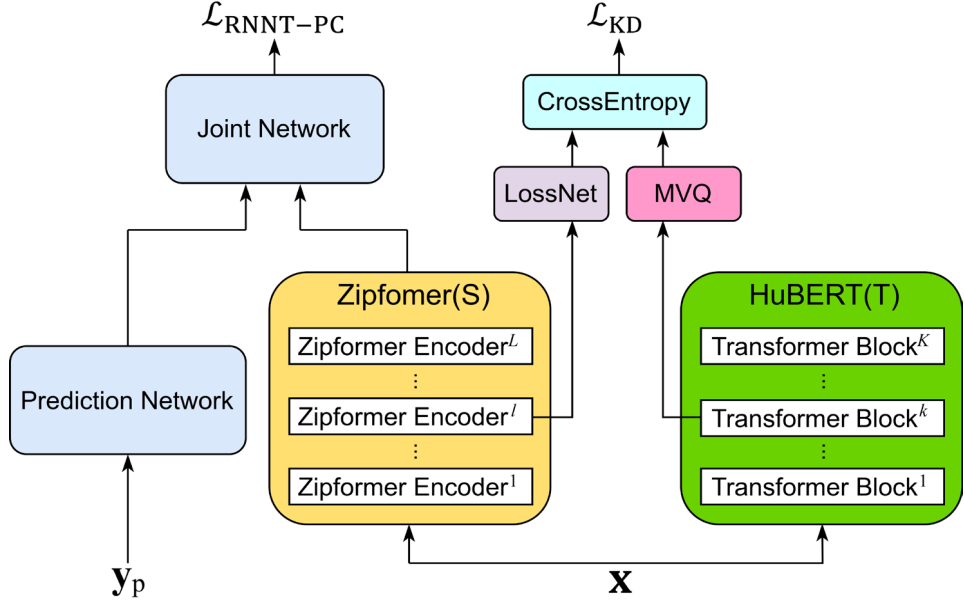

**Fig. 1.** The architecture of the proposed model. The student model is Zipformer, the acoustic encoder in RNN-T. The teacher model is a fine-tuned version of HuBERT. **x** is the input speech sequence, $\mathbf{y}_p$ is the sequence of previously predicted non-blank labels.

## II. Proposed Method

### A. RNN-T based ASR with Punctuation and Capitalization

In our approach, we utilize the neural transducer system, known as RNN-T, due to its simplicity and exceptional performance in streaming mode. E2E ASR can fundamentally be formulated as sequence-to-sequence problem. We denote the input speech utterance as $\mathbf{x} = \{\mathbf{x}_1 \ldots \mathbf{x}_T\}$, where $T$ is the number of frames. The wordpiece inventory, which includes punctuation marks and mixed-cased tokens, is denoted as $Y$. The corresponding label sequence formatted with **p**unctuation and **c**apitalization is represented as $\mathbf{y}_{pc} = \{y_1 \ldots y_U\}$, where $U$ is the length of label sequence and $y_u \in Y$.

We define $Y_b$ as $Y \cup \{\langle b \rangle\}$, where $\langle b \rangle$ represents the blank label. Denote $Y_b^*$ as the set of all sequences over output space $Y_b$, and the element $\mathbf{a} \in Y_b^*$ as an alignment sequence. The posterior probability is then expressed as:

$$P(\mathbf{y}_{pc}|\mathbf{x}) = \sum_{\mathbf{a} \in \mathcal{B}^{-1}(\mathbf{y}_{pc})} P(\mathbf{a}|\mathbf{x}) \quad (1)$$

where $\mathcal{B}: Y_b^* \to Y^*$ is a function that removes blank symbols from an alignment $\mathbf{a}$. The alignment probability $P(\mathbf{a}|\mathbf{x})$ is formulated under the assumption that the output at time t is conditionally dependent on the sequence of previous non-blank predictions:

$$P(\mathbf{a}|\mathbf{x}) = \prod_{\tau} P(a_{\tau}|\mathbf{p}, \mathbf{x}) \quad (2)$$

where $a_\tau$ represents the output at position $\tau$, **p** represents the non-blank outputs of the prediction network, and $P(a_\tau|\mathbf{p}, \mathbf{x})$ represents the probability of observing $a_\tau$ at position $\tau$. The RNN-T loss function is then the negative log of the posterior probability as in Eq. (1):

$$\mathcal{L}_{\text{RNNT-PC}} = -\log P(\mathbf{y}_{pc}|\mathbf{x}) \quad (3)$$

### B. MVQ-based Knowledge Distillation

Knowledge distillation (KD) is a frequently used approach to improve the performance of RNN-T based ASR systems [22, 25]. Unlike traditional KD, where the teacher labels are often distribution lattices [25, 26] or float type embeddings [27, 28], MVQ-KD compresses the embedding from an intermediate layer of teacher model to codebook indexes (CI). The student model is then trained to predict its CI. Denote $\mathbf{T}_t^k$ as the embedding extracted from the $k$-th layer of the teacher model and $\mathbf{S}_t^l$ as the embedding at $l$-th layer of the student model at $t$-th frame. Let $\mathbf{i}_t = (i_{t,1}, \ldots, i_{t,N})$ be the compressed CI of $\mathbf{T}_t^k$, where $N$ is the number of codebooks and $i_{t,n}$ indicates which entry in the $n$-th codebook is selected. The knowledge distillation loss function is:

$$\mathcal{L}_{\text{KD}} = \sum_{t=1}^{T} \sum_{n=1}^{N} \text{CrossEntropy}\left(\mathbf{i}_t, \text{LossNet}(\mathbf{S}_t^l)\right) \quad (4)$$

where LossNet is a neural network that converts the student embedding into probabilities through a linear layer with softmax activation. At each timestamp $t$, MVQ-KD conducts $N$ independent classifications. We adopt an 8-bit integer to represent each entry of $\mathbf{i}_t$. For instance, the sequence (24, 0, 1, 27, 255, 12, 17, 10) serves as an example of CI when $N$=8.

Given that the dimension of the teacher embeddings in this work is 512, a compression rate of 128 can be achieved if $N=16$ is used. By means of MVQ-KD, the teacher labels can be precomputed and saved to disk with a high compression rate so that the computation and storage issue in traditional KD can be significantly alleviated. The detailed configuration of the student model and the teacher model are provided in Sec. III.B

*C. Fusion with Auxiliary Loss*

We utilize Eq. (3) as the primary loss function, which is sufficient for training E2E formatted ASR model. Different from [11], where an auxiliary CTC loss derived from the token sequence without punctuation and the output of the intermediate layer is employed to facilitate the model training, we adopt knowledge distillation loss formulated in Eq. (4) as the auxiliary loss function. The fusion with the auxiliary loss is performed through a weighted combination of the two loss functions:

$$\mathcal{L} = \mathcal{L}_{\text{RNNT-PC}} + \alpha \mathcal{L}_{\text{KD}} \quad (5)$$

where $\alpha$ is a hyper-parameter that controls the distillation loss weight.

The fusion with the knowledge distillation loss enhances the student model by enabling it to learn the rich information from the intermediate layer of the teacher model. The architecture of the proposed model is illustrated in Fig. 1. During inference, the LossNet, as described in Sec. 2.2, is excluded, resulting in no overhead in inference time.

III. EXPERIMENT SETUP

*A. Dataset*

We utilized LibriSpeech-PC [15] dataset for training, which enhances the original transcriptions of the LibriSpeech corpus [30] by restoring punctuation and capitalization. During the restoration process, samples were excluded if they contained non-standard Unicode characters, significant text deviations, or durations under one second. After restoration process, over 90% of the 960 hours of audio data in the LibriSpeech corpus were retained. The evaluation is conducted on the test-clean and test-other subsets.

LibriSpeech-PC includes a wide variety of punctuation marks and word casing forms. In this experiment, we implement the following preprocessing steps for the transcripts: 1) We focus exclusively on periods, commas, and question marks, excluding other punctuation marks due to their limited presence in the dataset; 2) Words and punctuation marks are separated by a space; 3) Multiple consecutive whitespaces are consolidated into a single whitespace.

*B. Implementation Details*

The extra-large version of HuBERT [29], fine-tuned on the entire LibriSpeech dataset, is utilized as the teacher model[1]. The medium-scale version of Zipformer [20] is adopted as student model. The extra-large HuBERT model comprises a total of 48 layers. In accordance with the selection in [21], we extract the embeddings from the 36th transformer block to generate codebook indexes (CI), as this layer offers rich information while being less difficult to learn from. For the student model, following the classic setup in [20], the number of layers in its six stacks are {2, 2, 3, 4, 3, 2}. We perform KD on the 4th stack, specifically the 11th layer, of the student model to share the similar relative position as in the teacher model. The number of codebooks is set to 16, as recommended in [21]. The CI of the entire training set are generated using MVQ and stored to disk, which are retrieved when training.

We consider two fully formatted E2E ASR models: the **Whisper-large** model[2] and our **Zipformer-KD** model. The proposed model is developed using the icefall[3] framework. We utilize the default configuration from the icefall LibriSpeech recipe[4] to setup Zipformer. The 80-dimensional Mel filter bank features are used as the input acoustic features. The SentencePiece Unigram tokenizer [31] is adopted with a vocabulary of 500 tokens whose detokenized output includes upper- and lower-case characters, as well as commas, periods, and question marks. By default, we set $\alpha$ in Eq. (5) to 0.1. The model is trained on four 32GB NVIDIA Tesla V100 GPUs over 30 epochs, with the final model derived by averaging the last ten best checkpoints. The beam search of size 4 is employed for decoding.

*C. Training External Language Model*

During the decoding process, an external language model (LM) can be employed to improve the performance of E2E ASR systems. In this work, we utilize an external LM which is composed of a 3-layer long short-term memory (LSTM) network with 1024 units. It shares the same vocabularies as the transducer model and is trained on LibriSpeech-PC transcriptions over 10 epochs. The checkpoint from the final epoch is used for decoding. The outputs of the transducer model and the LM are fused through log-linear interpolation, also known as shallow fusion [32].

*D. Metrics*

To measure the accuracy of the predicted text with punctuation and capitalization, we employ two types of metrics. The first type is F1-based metrics, including precision, recall and F1-scores, which are commonly used in studies for punctuation and capitalization [33]. However, calculating the F1-score for all samples in ASR applications is not feasible, as the model's output may contain ASR errors. In this work, we compute F1-based metrics only for samples where WER is zero. The second type is WER-based metrics, including classic word error rate with no punctuation and no capitalization (WER), word error rate with capitalization and no punctuation (WER C), word error rate with both punctuation and capitalization (WER PC) and punctuation error rate (PER) [15].

IV. EXPERIMENT RESULTS

*A. Comparison with State-of-the-art Models*

The results of the F1-based metrics are illustrated in Table 1. In the test-clean subset of 2417 samples, the E2E Conformer [15] achieves a zero WER in 66% of samples, whereas Zipformer-KD improves this to 76%, demonstrating enhanced error-free output capability. For punctuation, the E2E Conformer records an F1-score of 87.40, while Zipformer-KD enhances this to 88.57, indicating superior punctuation prediction. For capitalization, Zipformer-KD exceeds the E2E Conformer with an F1-score of 95.76 compared to 94.85, thereby exhibiting consistent performance enhancement. In the test-other subset of 2856 samples, Zipformer-KD outperforms with a zero WER in 59% of samples versus the E2E Conformer's 50%. Furthermore, Zipformer-KD's punctuation F1-score rises to 89.03, surpassing the E2E Conformer's 88.27. These results suggest that Zipformer-KD consistently edges out in error reduction and accuracy, making it potentially more effective for enhancing formatted text accuracy in varied acoustic environments. To ensure a fair comparison of the models, it is imperative to consider only the intersection of the zero-WER sample subsets. Nonetheless, the E2E Conformer [15] does not provide explicit data for its zero-WER samples. Consequently, we are constrained to present the results based solely on the respective zero-WER sample subsets of each model.

---

[1] https://dl.fbaipublicfiles.com/hubert/hubert_xtralarge_ll60k_finetune_ls960.pt
[2] https://huggingface.co/openai/whisper-large
[3] https://github.com/k2-fsa/icefall
[4] https://github.com/k2-fsa/icefall/blob/master/egs/librispeech/ASR/zipformer/zipformer.py

**Table 1.** Number of samples where WER is zero and their F1-score, Precision and Recall for Punctuation and Capitalization.

| Model | Subset (Total samples) | WER=0 samples | Punctuation | | | Capitalization | | |
|---|---|---|---|---|---|---|---|---|
| | | | F1 | Prec. | Recall | F1 | Prec. | Recall |
| E2E Conformer [15] | test-clean (2417) | 1605(66%) | 87.40 | 86.11 | 88.90 | 94.85 | 95.53 | 94.20 |
| Zipformer-KD (ours) | | 1849(76%) | 88.57 | 88.10 | 89.05 | 95.76 | 95.80 | 95.73 |
| E2E Conformer [15] | test-other (2856) | 1439(50%) | 88.27 | 86.87 | 89.93 | 95.50 | 96.37 | 94.67 |
| Zipformer-KD (ours) | | 1704(59%) | 89.03 | 88.23 | 89.84 | 96.13 | 96.73 | 95.53 |

**Table 2.** Results of WER-based metrics on LibriSpeech-PC test-clean and test-other sets.

| Model | Num. Param. | test-clean (%) | | | | test-other (%) | | | |
|---|---|---|---|---|---|---|---|---|---|
| | | PER | WER | WER C | WER PC | PER | WER | WER C | WER PC |
| Whisper-base [15] | 74M | 36.97 | 4.50 | 6.82 | 11.23 | 40.27 | 11.23 | 13.57 | 17.82 |
| E2E Conformer [15] | 115M | 29.48 | 2.83 | 4.72 | 8.11 | 27.37 | 5.65 | 7.57 | 10.48 |
| Whisper-large | 1550 M | 30.52 | 3.02 | 5.15 | 8.60 | 30.55 | 5.99 | 8.02 | 11.29 |
| Zipformer-KD (ours) | 65M | 29.74 | 2.20 | 4.20 | 7.58 | 28.18 | 5.23 | 7.27 | 10.21 |
| E2E Conformer+LM [15] | 115M | 29.24 | 2.22 | 4.05 | 7.66 | 27.67 | 4.67 | 6.55 | 9.80 |
| Zipformer-KD+LM (ours) | 65+25M | **27.24** | **1.87** | **3.59** | **6.65** | **26.55** | **4.56** | **6.28** | **9.05** |

Table 2 presents the results of WER-based metrics on LibriSpeech-PC test-clean and test-other subsets. In the absence of an external language model, Zipformer-KD, with 65M parameters, exhibits competitive performance in WER PC, achieving scores of 7.58% on the test-clean subset and 10.21% on the test-other subset. This demonstrates its exceptional capability to accurately handle text intricacies like punctuation and capitalization, outperforming both Whisper variants. In terms of PER, Zipformer-KD achieves scores of 29.74% on the test-clean subset and 28.18% on the test-other subset, which are competitive with larger models like the E2E Conformer [15]. It is noteworthy that Zipformer-KD achieves lower WER and WER PC than the E2E Conformer+LM [15] on the test-clean subset, despite not utilizing a language model. With the integration of an external language model, the Zipformer-KD+LM model further enhances its performance, achieving the best overall performance. It records WER PC scores of 6.65% on the test-clean subset and 9.05% on the test-other subset, representing a significant relative improvement of approximately 13% and 8%, respectively, over previous state-of-the-art models. This improvement is achieved while maintaining a smaller parameter size. Additionally, the model's PER is reduced to 27.24% for the test-clean subset and 26.55% for the test-other subset, underscoring its refined accuracy and efficiency.

*B. Ablation Studies*

We perform ablation studies for knowledge distillation (KD) in both non-streaming and streaming modes. As reported in Table 3, in non-streaming mode, the absence of KD results in degradations in all metrics, with WER PC increasing by 0.24% and 0.49% and PER increasing by 0.49% and 1.33% on the test-clean and test-other subsets respectively. Conversely, the implementation of KD significantly improves these metrics, suggesting the advantage of using KD. In streaming mode, the impact of KD is further pronounced, reducing WER PC by 0.47% and 0.15%, and PER by 1.27% and 0.24% on the test-clean and test-other subsets respectively. These results manifest the effectiveness and robustness of KD in enhancing fully formatted ASR performance. Given that the teacher model is the HuBERT model fine-tuned on the LibriSpeech dataset where the transcripts are normalized texts, incorporating KD guides the intermediate representations of the student model to be more conducive to the accurate prediction of normalized text. Notably, punctuation marks in the output are highly sensitive to the correctness of the previously predicted texts. Within the RNN-T decoder, the prediction of subsequent tokens relies on prior output tokens; thus, the accuracy of previously predicted text can significantly enhance the performance of punctuation mark prediction. Moreover, capitalized tokens typically appear alongside punctuation marks; for instance, the word following a terminal period is always capitalized. Therefore, the improvements in punctuation prediction also facilitates more accurate prediction of capitalized tokens. Hence, the application of KD delivers notable gains in fully formatted ASR performance.

**Table 3.** Ablation studies for KD on LibriSpeech-PC dataset in non-streaming and streaming modes. PER, WER, WER PC (in brackets) and training cost are reported. Training cost is measured in total GPU hours.

| Model | test-clean (%) | | test-other (%) | | GPU Hours |
|---|---|---|---|---|---|
| | PER | WER(PC) | PER | WER(PC) | |
| **Non-streaming Models** | | | | | |
| Zipformer w/o KD | 30.23 | 2.42 (7.82) | 29.51 | 5.55 (10.70) | 152 |
| Zipformer w/ KD | 29.74 | 2.20 (7.58) | 28.18 | 5.23 (10.21) | 164 |
| **Streaming Models** | | | | | |
| Zipformer w/o KD | 30.01 | 3.13 (9.00) | 29.05 | 7.85 (12.90) | 188 |
| Zipformer w/ KD | 28.74 | 2.85 (8.53) | 28.81 | 7.55 (12.75) | 200 |

**Table 4.** Ablation studies on tuning hyper-parameter $\alpha$.

| Hyper-parameter | WER PC (%) | |
|---|---|---|
| | test-clean | test-other |
| $\alpha = 0.05$ | 7.62 | 10.21 |
| $\alpha = 0.1$ (**final**) | **7.58** | **10.21** |
| $\alpha = 0.2$ | 7.67 | 10.22 |

**Table 5.** An example of model outputs with and without KD.

| | |
|---|---|
| Gold | Who is Humpy Dumpy? asked the Mice. |
| Zipformer w/o KD | Who is Uncy Dumpy? asked the Mice. |
| Zipformer w/ KD | Who is Humpy Dumpy? asked the Mice. |

Regarding training cost, the introduction of KD results in only a slight increase in total GPU hours - an increment of 12 hours in both the non-streaming and streaming modes. Since the teacher labels, i.e. the codebook indexes, are prepared prior to training, there is no need for computationally intensive on-the-fly generation of teacher labels during training. Consequently, the limited overhead is primarily due to the additional steps of reading the codebook indexes from local storage and calculating the KD loss, as described in Eq. (4).

Table 4 presents the results of tuning hyper-parameter $\alpha$ in Eq. (5). We tune the distillation loss weight $\alpha$ over three different values {0.05, 0.1, 0.2}, while keeping all other training parameters unchanged. The optimal performance on both the test-clean and test-other subsets, as reported in Table 4, is achieved when $\alpha$ is set to 0.1.

*C. Example Output*

Table 5 presents an example of model outputs with and without knowledge distillation (KD). Both models exhibit proficiency in predicting punctuation and capitalization. However, the model trained without KD struggles to accurately recognize less frequent words, whereas the implementation of KD enables correct recognition of these less frequent words.

V. CONCLUSION

In this work, we introduce an enhanced fully formatted E2E ASR model that employs knowledge distillation through multi-codebook vector quantization. A transducer system, utilizing the high-performance Zipformer as the acoustic encoder, is constructed and trained on formatted text. Intermediate layer embeddings from a self-supervised, pre-trained teacher model are compressed to codebook indexes (CI). The student model, specifically the acoustic encoder, is trained to predict its CI. Experimental results on LibriSpeech-PC datasets demonstrate the superior performance of the proposed model. Additionally, ablation studies validate the effectiveness of knowledge distillation in enhancing fully formatted ASR performance.